\documentclass[sigconf]{acmart}

\copyrightyear{2021}
\acmYear{2021}
\setcopyright{acmcopyright}
\acmConference[SIGMOD '21]{Proceedings of the 2021
International Conference on Management of Data}{June 20--25, 2021}{Virtual Event,
China}
\acmBooktitle{Proceedings of the 2021 International Conference on Management of
Data (SIGMOD '21), June 20--25, 2021, Virtual Event, China}
\acmPrice{15.00}
\acmDOI{10.1145/3448016.3457333}
\acmISBN{978-1-4503-8343-1/21/06}

\usepackage[utf8]{inputenc} 
\usepackage[T1]{fontenc}    
\usepackage{hyperref}       
\usepackage{url}            
\usepackage{booktabs}       
\usepackage{amsfonts}       
\usepackage{nicefrac}       
\usepackage{microtype}      
\usepackage{float}
\usepackage{algorithmic}
\usepackage{algorithm, setspace}
\usepackage{graphicx}
\usepackage{amsmath}
\usepackage{amsfonts}
\usepackage{svg}
\usepackage{subcaption}
\usepackage{soul}
\usepackage{ulem}
\usepackage{xcolor}
\usepackage{balance}

\begin{document}
\fancyhead{}
\title{Fast Processing and Querying of 170TB of Genomics Data via a Repeated And Merged BloOm Filter (RAMBO)}

\author{Gaurav Gupta}
\authornote{Equal contribution.\\ Code available at: \hyperlink{https://github.com/gaurav16gupta/RAMBO_MSMT}{https://github.com/gaurav16gupta/RAMBO\_MSMT}}
\affiliation{Electrical and Computer Engineering Department, Rice University}
\email{gaurav.gupta@rice.edu }

\author{Minghao Yan}
\authornotemark[1]
\affiliation{Department of Computer Science, Rice University}
\email{my29@rice.edu}

\author{Benjamin Coleman}
\affiliation{Electrical and Computer Engineering Department, Rice University}
\email{ben.coleman@rice.edu}
\author{Bryce Kille }
\affiliation{Department of Computer Science, Rice University}
\email{brycekille@gmail.com}

\author{R. A. Leo Elworth}
\affiliation{Department of Computer Science, Rice University}
\email{chilleo@gmail.com}

\author{Tharun Medini }
\affiliation{Electrical and Computer Engineering Department, Rice University}
\email{tharun.medini@rice.edu}

\author{ Todd Treangen }
\affiliation{Department of Computer Science, Rice University}
\email{treangen@rice.edu}

\author{Anshumali Shrivastava }
\affiliation{Department of Computer Science, Rice University}
\email{anshumali@rice.edu}


\renewcommand{\shortauthors}{Gupta et al.}


\begin{abstract}
DNA sequencing, especially of microbial genomes and metagenomes, has been at the core of recent research advances in large-scale comparative genomics. The data deluge has resulted in exponential growth in genomic datasets over the past years and has shown no sign of slowing down. Several recent attempts have been made to tame the computational burden of sequence search on these terabyte and petabyte-scale datasets, including raw reads and assembled genomes.
However, no known implementation provides both fast query and construction time, keeps the low false-positive requirement, and offers cheap storage of the data structure.\\
We propose a data structure for search called RAMBO (Repeated And Merged BloOm Filter) which is significantly faster in query time than state-of-the-art genome indexing methods- COBS (Compact bit-sliced signature index), Sequence Bloom Trees, HowDeSBT, and SSBT. Furthermore, it supports insertion and query process parallelism, cheap updates for streaming inputs, has a zero false-negative rate, a low false-positive rate, and a small index size.
RAMBO converts the search problem into set membership testing among $K$ documents. Interestingly, it is a count-min sketch type arrangement of a membership testing utility (Bloom Filter in our case). The simplicity of the algorithm and embarrassingly parallel architecture allows us to stream and index a 170TB whole-genome sequence dataset in a mere 9 hours on a cluster of 100 nodes while competing methods require weeks.  
\end{abstract}

\ccsdesc[500]{Information systems~Search engine indexing}
\ccsdesc[100]{Information systems~Distributed retrieval}
\ccsdesc[500]{Theory of computation~Bloom filters and hashing}
\ccsdesc[300]{Applied computing~Computational genomics}

\keywords{Information retrieval; Genomic sequence search; Bloom filter}

\maketitle
\section{Introduction}

\begin{table*}[ht]
\centering
\fontsize{9}{11}\selectfont

\begin{tabular}{|c|c|c|c|}
\hline
Method & Size & Query Time & Comments \\
\hline
Inverted Index & Best case:  & $O(1)$ & Enormous construction time, Impractical for bigger datasets \\
& $\log K \bigcup_{S\in\mathcal{S}}|S|$ & & Best case needs MPH and a known k-mer (term) distribution \\ 
\hline
BIGSI/COBS & $\sum_{S\in\mathcal{S}}|S|$ & $O(K)$ & Query time is linear in $K$, Small size index\\ \hline
Sequence Bloom Trees & $\log K \sum_{S\in\mathcal{S}}|S| $ & Best: $O(logK)$, Worst: $O(K)$   & Sequential query process is bottleneck \\
\hline
RAMBO  & $\Gamma \log K  \sum_{S \in \mathcal{S}} |S| $ & $O(\sqrt{K} \log K)$ & $\Gamma <1$, Sub-linear query time\\
\hline
\end{tabular}
\vspace{0.2cm}
\caption{Theoretical comparison of related algorithms on sequence searching. $S \in \mathcal{S}$ represents a document. $K$ is the total number of documents. Here $\sum_{S\in\mathcal{S}}|S|$ represents total number of terms in $K$ documents and $\bigcup_{S\in\mathcal{S}}|S|$ is total unique terms.  MPH is minimal Perfect Hashing \cite{botelho2007simple}. For the Inverted Index size, the extra $\log K$ comes from the bit precision document IDs. For SBTs, $\log K$ is the height of the tree and Bloom Filters at each level is $O(\sum_{S\in\mathcal{S}}|S|)$ big in total. Refer to Section \ref{TheroryAnalysis} for the detailed analysis of RAMBO.}
\label{tab:relatedwork}
\vspace{-0.5cm}
\end{table*} 

The availability of genomic data facilitates necessary biological research like cancer genomics, vaccine development and immunization, infection tracking, early diagnosis and treatments, structural variant analyses, and more \cite{stevens2017public} \cite{snitkin2012tracking}. Recent advances in DNA sequencing technologies have both increased the throughput and decreased the cost of reading the DNA sequence of organisms and microbial communities of interest. While this has broadened the horizons of biological research, it poses new challenges for computational biologists. Thanks to these advancements, genome sequence data has doubled in size every 2 years and is likely to grow at an increasing pace~\cite{b3, schatz2013dna}. The European Nucleotide Archive (ENA) and NCBI Archive already contain petabytes of data. It has become computationally prohibitive to search these vast archives for DNA sequences of interest. Efficient and frugal search functionality across all available genomic and metagenomic datasets is significant to public health. It would enable quick identification of already-sequenced organisms that are highly similar to an outbreak strain. 
 
\begin{figure}
    \centering
    \includegraphics[scale=0.4]{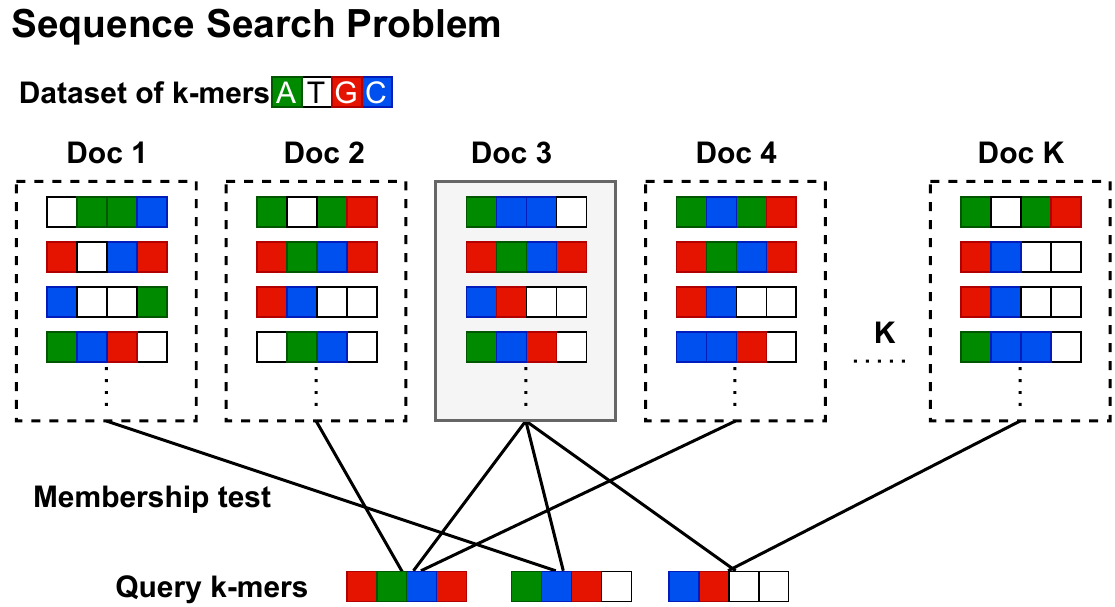}
    \caption{Sequence search problem: First, we convert each of the $K$ documents into a set of k-mers. The k-mers of length $31$ are generated using a sliding window on the sequence (k=4 in the figure for illustration). Given the k-mers from a query sequence, the task is to determine which of the $K$ documents contain all the k-mers present in the query. }
    \vspace{-0.6cm}
    \label{fig:problemSc}
\end{figure}

The DNA sequence search problem is analogous to Document Retrieval. Given a query gene strand, we are expected to retrieve the whole gene sequence that contains it (Figure \ref{fig:problemSc}). The search results are critical for a variety of genomic research and analysis tasks.
Its similarities with the problem of web search, in terms of both objective and scale, have triggered a flurry of ideas borrowed from the information retrieval community \cite{ondov2016mash, crainiceanu2013bloofi}. In the seminal work BLAST~\cite{altschul1990basic}, a popular search platform for biological databases, the authors provided the first attempt to search over large databases. However, the method does not scale to large query datasets \cite{altschul1990basic} due to the reliance on computationally expensive local sequence alignment. On the other hand, traditional approaches such as the inverted index~\cite{croft2010search} cannot quickly index large-scale data without violating memory constraints.

To address this issue, computational biologists and database practitioners have shifted their attention to Bloom Filter based methods~\cite{bloom1970space, b3, Bingmann2019COBSAC} and similar bit-signature approaches for gene sequence search due to the sheer scale of genomic data. 
A recent \textit{Nature Biotechnology} article, \textbf{BIGSI} \cite{b3}, proposed a method which was successful in indexing the set of 469,654 bacterial, viral and parasitic DNA sequences in the European Nucleotide Archive \cite{b12} of December 2016. These sequences come from read archive datasets (FASTQ format) or assembled genomes (FASTA format) consisting of raw sequences from a DNA sequencer or genome assembler, respectively. The average length of each of these half-million sequences is more than 100M characters.
This makes the entire archive database about 170TB in size. 

To create the index, BIGSI and many other practical indices convert a long gene sequence into a set of length-$31$ strings (each shifted by 1 character) and compress the strings using a Bloom Filter (sometimes called a Bitsliced signature). These length-$31$ strings are called $k$-mers. It is analogous to "terms" in the information retrieval literature. Specifically, a $k$-mer is a character $n$-gram where $k=n$. In our experiments $k=31$, just like most of the state of the art methods. We explain the rationale behind this choice in Section \ref{Experiments}.
BIGSI essentially creates a Bloom Filter for each document (a set of $k$-mers for one microbe). This index is simply an array of independent Bloom Filters where the query time grows \textit{linearly} in the number of documents.
The \textbf{Sequence Bloom Tree (SBT)} \cite{SBT, crainiceanu2013bloofi} is another approach to solve the sequence search problem on the scale of the entire sequence read archive (SRA)~\cite{kodama2011sequence} using Bloom Filters. To achieve sublinear query time complexity, the SBT uses a tree-like hierarchy of Bloom Filters~\cite{SBT}. However, this introduces a substantial memory overhead at each node. Moreover, the query process cannot enjoy the parallelism of BIGSI because tree-based traversal is a sequential algorithm. Experimental results from~\cite{b3} suggest that SBTs become less scalable when the time and evolution of species are factored in, which is the case for bacteria and viruses. Several follow-ups using ideas similar to SBTs also suffer from the same issues \cite{b5, b6, b7}. 
By removing the hierarchy, BIGSI and its recent follow-up \textbf{COBS (Compact bit-sliced signature index)} \cite{Bingmann2019COBSAC} obtained substantial memory savings compared to SBTs. The simplicity of the index, combined with embarrassingly parallel architecture and clever bit manipulation tricks, enables BIGSI and COBS to process and search over 170TB WGS datasets in a record query time. However, with an exponential increase in the number of datasets in the sequence archive, the linear scaling of latency and energy is too expensive.\\
\noindent{\bf Our Focus:} We propose methods to reduce the query cost of sequence search over the archive of dataset files to address the sheer scale and explosive increase of new sequence files. In particular, unlike BIGSI and COBS, we do not want the number of Bloom Filters used in the query to be of the same order as the number of datasets, which can run into several million. At the same time, we also want an algorithm that maintains all other beneficial BIGSI and COBS features. We are looking for a data structure for sequence search, which has the following properties: \textbf{1.} A zero false-negative rate, \textbf{2.} A low false-positive rate, \textbf{3.} Cheap updates for streaming inputs, \textbf{4.} Fast query time, and \textbf{5.} A simple, system-friendly data structure that is straightforward to parallelize. The system should have all these properties with the least possible memory size. \\

\vspace{-5pt}
\noindent{\bf Insights from Computer Science Literature:} There is a fundamental algorithmic barrier at the heart of this problem. The classical sub-linear search data structure provides tree-based solutions that mainly implement the SBT \cite{SBT}. However, trees complicate the query process and have issues with balanced partitions, especially when dimensionality blows up. Fortunately, the Count-Min Sketch (CMS) Algorithm~\cite{Cormode2005CMS} from the data streaming literature provides a workaround. Our proposal for sequence search, Repeated And Merged BloOm Filter (RAMBO) is a CMS using Bloom Filters.
It is a simple and intuitive way of creating merges and repetitions of Bloom Filters for membership testing over many sets. RAMBO leads to a better query-time and memory trade-off in practice. It beats the current baselines by achieving a very robust, low memory and ultrafast indexing data structure.
\vspace{-0.2cm}
\subsection{Contribution}
\label{contribution}
Instead of having separate Bloom Filters for each document, we split the documents into a small number of random partitions. We keep one Bloom Filter for each partition. Inspired by the theory of the Count-Min Sketch~\cite{Cormode2005CMS}, if we repeat the partitioning process with different random seeds a small number of times, we can return documents with high accuracy. 

Our proposed index RAMBO leads to massive improvements in query cost and would allow effortless scaling to millions of documents. We provide a rigorous experimental and theoretical analysis of the query time. Experimental comparisons show that RAMBO is significantly faster (between \textbf{25x} to \textbf{2000x} improvement) in query time over the most competitive baselines of gene data indexing while preserving competitive false-positive rates.

RAMBO can be made embarrassingly parallel for insertion and query by an intelligent choice of hash functions and judicious utilization of parallel processors. It can be easily distributed over multiple nodes, cores, and threads for achieving substantial speedup. We show remarkable improvements in construction and query time over the baselines on a 170TB WGS dataset. We reduce the time of offline construction of the index from 6 weeks (1008 hours for BIGSI) to only 9 hours (including the additional download time of 8 hrs). This is attributed to the fact that BIGSI downloads and indexes 460,500 files (170TB) sequentially. Its successor, COBS \cite{Bingmann2019COBSAC}, is much faster and better in all aspects than BIGSI. The RAMBO index has a slightly larger memory requirement than an optimal array of Bloom Filters (COBS), but it keeps a cheap index (1.8 terabytes) for 170TB worth of data. It is important to note that Bloom Filters in RAMBO can be replaced with any other set membership testing method.

\section{Preliminaries}

\subsection{Bloom Filters}
\label{BFintro}
The Bloom Filter ~\cite{bloom1970space,mitzenmacher2002compressed, cohen2003spectral} is an array of $m$ bits which represents a set $S$ of $n$ elements. It is initialized with all bits set to 0. During construction, we apply $\eta$ universal hash \cite{carter1978exact} functions $\{h_1, h_2...h_{\eta}\}$ with range $m$ to the elements of $S$ ($\eta$ and $n$ are different). We set the bits at the respective locations $\{h_1(x), h_2(x)...h_{\eta}(x)\}$ for each key $x \in S$. 
Once the construction is done, the Bloom Filter can be used to determine whether a query $q\in S$ by calculating the AND of the bits at the $\eta$ locations: $h_1(q), h_2(q)...h_{\eta}(q)$. The output will be $\mathrm{True}$ if all $\eta$ locations are $1$ and $\mathrm{False}$ otherwise. 
Bloom Filters have no false negatives as every key $x \in S$ will set all the bits at locations $\{h_1(x), h_2(x)...h_{\eta}(x)\}$. However, there are false positives introduced by hash collisions.
The false positive rate of the Bloom Filter, $p$, is given by: $  p = \left(1-\left[1-{\frac {1}{m}}\right]^{\eta n}\right)^{\eta} \approx \left(1-e^{-\eta n/m}\right)^{\eta} $.
We should note that this expression makes many simplifying assumptions and is not entirely correct, as we assume independence of the probabilities of each bit being set. A more accurate analysis is given in Christensen et al \cite{CHRISTENSEN2010944}. However, its deviation from practical numbers is minimal when the Bloom Filter size is large (Figure 2 of \cite{CHRISTENSEN2010944}). At the scale that we are dealing with, the difference becomes insignificant.
 
Using the simplified analysis, the false positive rate is minimized when we use $\eta = -\frac{\log p}{\log 2}$ and $m = -n \frac{\log p}{\log 2}$.
The size of a Bloom Filter grows linearly in the cardinality $n$ of the set it represents. Bloom Filter has a constant-time query operation.
\vspace{-0.2cm}
\begin{figure}[h]
    \centering
    \includegraphics[scale=0.3]{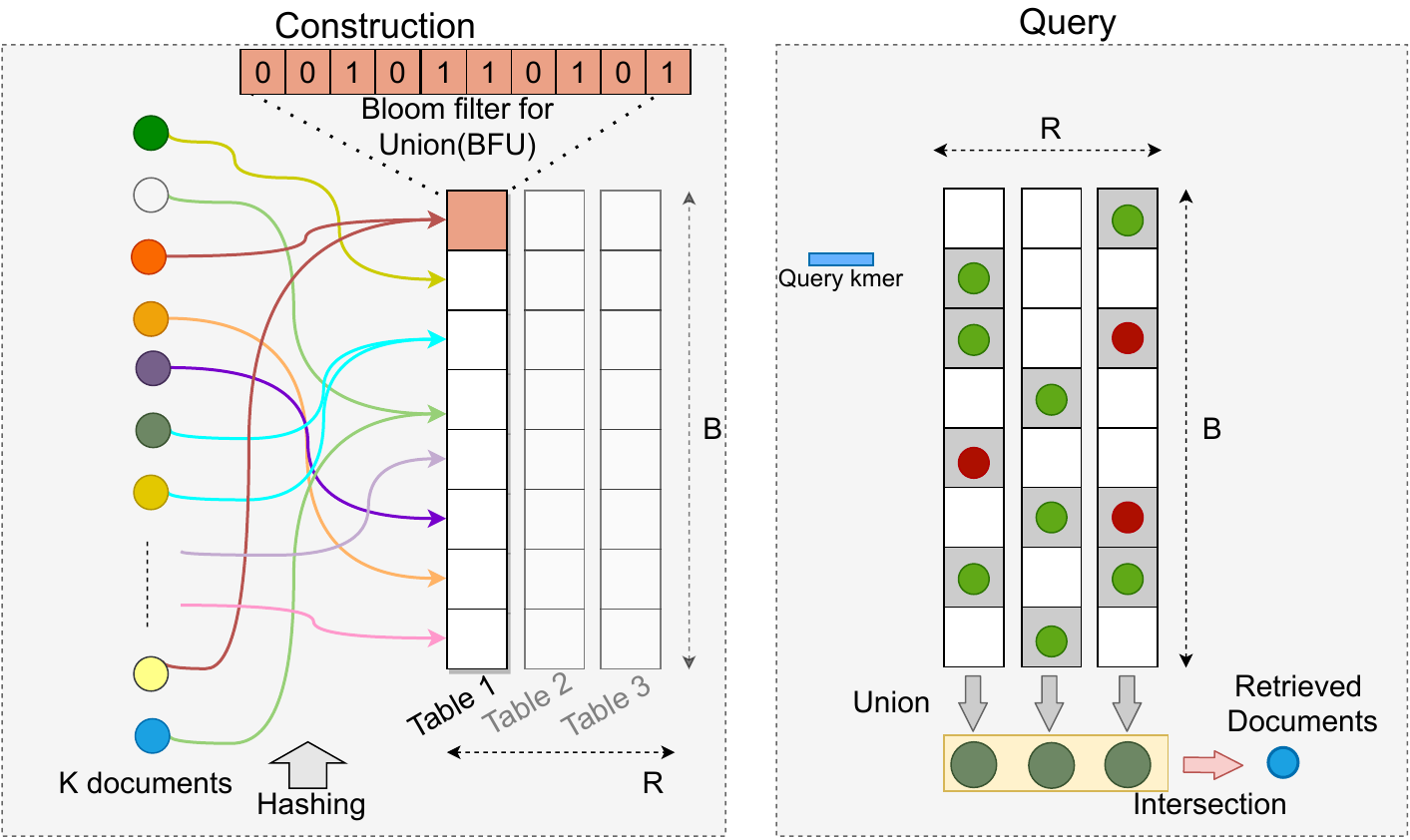}
    \caption{(a) Left: RAMBO architecture and the insertion process. The construction of the first repetition is highlighted. Here the $K$ documents are randomly partitioned (via a 2-universal hash function \cite{carter1978exact}). Each Bloom Filter (called BFU) is the union of sets equivalent of partitioned documents. (b) Right: For a given query each table, RAMBO returns one or more BFUs (represented by the dots) where the membership is defined. The red dot represents the false positives and green dot represents the true positives. The membership of a query $k$-mer is defined by the union of the returned documents from each repetition followed by the intersection across $R$ repetitions.
    }
    \label{fig:RAMBOarch}
\end{figure}

\vspace{-0.4cm}
\section{RAMBO: Repeated And Merged Bloom Filters}
\label{RAMBO}
The RAMBO architecture (Figure \ref{fig:RAMBOarch}) comprises an array of $R$ tables, each containing $B$ Bloom Filters. 
We partition the $K$ documents into $B$ groups and compress every group of documents to a Bloom Filter called Bloom Filters for the Union (BFU). 
Here each document is a set of $k$-mers.
This process is repeated independently for the $R$ tables using $R$ different and independent 2-universal hash functions. Due to the hash functions' universality, every cell of a table in RAMBO contains $K/B$ sets from $\mathcal{S}$ in expectation.

\vspace{-0.2cm}
\subsection{Intuition}
We have $K$ documents, partitioned into $B$ partitions, where $2 \le B \ll K$. Now, we are given a query term $q$. For simplicity, assume that this query belongs to a single document. Now, if we query each partition, we can determine which one of them contains $q$. We refer to this partition as $A_1$. Thus, with only $B$ Bloom Filter queries, we have reduced the number of candidate sets from $K$ to $\frac{K}{B}$ in expectation. If we independently repeat this process again, we find another partition $A_2$ that contains $q$. Our pool of candidate sets is now the set intersection of $A_1$ and $A_2$, which in expectation has size $\frac{K}{B^2}$. With more repetitions, we progressively rule out more and more options until we are left with only the sets that contain $q$. The critical insight is that each repetition reduces the number of candidates by a factor of $\frac{1}{B}$, which decreases \textit{exponentially} with the number of repetitions.
Since RAMBO is an extension of the Count-Min Sketch (CMS) data structure \cite{Cormode2005CMS}, most theoretical guarantees carry forward. We replace the counters in the CMS with Bloom Filters. Instead of adding counters to construct the CMS, we merge the sets of k-mer terms. The querying procedure of the CMS is replaced with an intersection over the merged sets to determine which sets contain a query term. 
\vspace{-0.3cm}
\subsection{Construction}
\vspace{-0.1cm}
We assign sets based on a partition hash function $\phi(.)$ that maps the set identity to one of $B$ cells. We use $R$ independent partition hash functions $\{ \phi_1, \ \phi_2, ..., \phi_R\}$. 
Suppose we want to add a set of terms in Doc-1 to RAMBO. We first use the partition function $\phi_i($Doc-1$)$ to map Doc-1 in repetition $i$, $\forall \ i \in \{0,\ 1, ..., R \}$. Then we insert the terms ($k$-mers) of Doc-1 in $R$ assigned BFUs (Algorithm \ref{alg:construction}). The Bloom Filter insertion process is defined in Section \ref{BFintro}. We define the size of each BFU based on the expected number of insertions in it. This is further analyzed in Section \ref{paramSelection}.

Our RAMBO data structure is a $B \times R$ CMS of Bloom Filters. Clearly, this structure is conducive to updates to a data stream. Every new term in a document is hashed to unique locations in $R$ tables. The size of the BFU can be predefined or a scalable Bloom Filter \cite{scalableBF} can be used for adaptive size. 
\begin{algorithm}
\begin{algorithmic}
\STATE {\bf Input:} Set $\mathcal{S}$ of $K$ sets 
\STATE {\bf Result:} RAMBO (size: $B \times R$ )
\STATE Generate $R$ partition hash functions $\phi_1(\cdot), ... \phi_R(\cdot)$
\STATE RAMBO $\leftarrow B\times R$ array of Bloom Filters
\WHILE{Input $S_i$}
    \FOR{term $x \in $ $S_i$}
        \FOR{$d = 1, ... R$}
            \STATE Insert($x$, RAMBO[$ \phi_d(x),d$])
        \ENDFOR
    \ENDFOR 
\ENDWHILE
 \end{algorithmic}
  \caption{Algorithm for insertion in RAMBO architecture}
  \label{alg:construction}
\end{algorithm}

\vspace{-0.3cm}
\subsection{Query}
\label{queryMethod} 
We start the RAMBO query process by performing membership testing for each of the $B \times R$ BFUs. This is followed by taking the union of the sets corresponding to each filter that returns True in each table, and then the intersection of those unions across the $R$ tables. The union and intersection are implemented using fast bitwise operations, and the expected query time is sub-linear (Section \ref{queryTime}). Algorithm \ref{alg:membershipTest} presents the query process. Here, set $A\subseteq \mathcal{S}$ is the final set of matched documents. 

Note that a $k$-mer can occur in multiple documents, which we call multiplicity. For this reason, and the fact that Bloom Filters have a nonzero false positive rate, multiple BFUs may return True (Figure \ref{fig:RAMBOarch}). If a query $k$-mer does not exist in any document, RAMBO will most likely return an empty set $A = \emptyset$ or a small set of false positives with low probability.

\subsubsection{Large Sequence Query}
\label{Fullsequencequery}
To query a larger term sequence with length $n$, we simply use a sliding window of size $k$ to go through the entire sequence. This will create a set of terms $Q$ to query. Then we iterate over the terms in $Q$ and membership test each term. The final output should be the intersection of all returned outputs from each term in $Q$.
Since Bloom Filter does not have any false negatives, we are guaranteed to obtain a valid result. We only need to perform exponentially less (in the cardinality of $Q$) number of membership tests as the first returned FALSE will be conclusive. It is interesting to note that the final output size is upper bounded by the output size of the rarest k-mer in the query sequence.
\begin{algorithm}
\begin{algorithmic}
\STATE {\bf Input:} query $q \in \Omega$
\STATE {\bf Architecture:} RAMBO ($M$) // Size $B \times R$ array of Bloom Filters. 
\STATE {\bf Result:} $A\subseteq \mathcal{S}$, where $q\in S_i$ $\forall$ $S_i \in A$
\STATE $Q = $Terms$(q)$
\FOR{$r=1:R$}
    \STATE $G_r = \{Null\}$
    \FOR {$b=1:B$}
    \STATE  $G_r = G_r \cup DocIDs(M[b,r])\ \textbf{if}\ Q \in $BFU$(M[b,r])$
    \ENDFOR
\ENDFOR
\STATE $A = \cap_{i} G_i$ \COMMENT{final returned Doc ID's} \\

\STATE {\bf Define:} Q\ $\in \ $BFU(M$[b,r]$)
  \STATE \  \  return True if $x \in $BFU(M$[b,r]$) $\forall x \in $Q
 \end{algorithmic}
 \caption{Algorithm for query using RAMBO architecture}
 \label{alg:membershipTest}
\end{algorithm}

\vspace{-0.4cm}
\section{Analysis}
\label{TheroryAnalysis}
\textbf{Problem Definition:}
We are given a set of $K$ documents $\mathcal{S} = \{ S_1, \ S_2, ..., \ S_K \}$. Each document $S_i$ contains $k$-mers from a universe $\Omega$ of all possible $k$-mers. Given a query $q \in \Omega$, our goal is to identify all the documents in $\mathcal{S}$ that contain $q$. That is, the task is to return the subset $A_q \subseteq \mathcal{S}$, such that $q \in S_i$ if and only if $S_i \in A_q$. 

RAMBO has two important parameters, $R$ and $B$, that control the resource-accuracy trade-off. In this section, we will analyze the false positive rate, query time, and index size to find the optimal values of R and B. 

\subsection{False-Positives}
\label{FPanal}
Our first claim is that RAMBO cannot report false negatives. This follows trivially from our merging procedure and the fact that each BFU cannot produce false negatives \cite{bloom1970space}. Next, we begin by finding the false positive rate of one document and extend this result to all $K$ documents.

\begin{lemma} \textbf{Per document False Positive Rate }\\
Given the RAMBO data structure with $B\times R$ BFUs, each with false positive rate $p$ and query $q$, we assume that $q$ belongs to no more than $V$ documents. Under these assumptions, the probability of incorrectly reporting that $q \in S_i$ when $q \not \in S_i$ is 
$$F_p = \left(p \left(1 - \frac{1}{B}\right)^V + 1 - \left(1 - \frac{1}{B}\right)^V\right)^R $$ 
where $p$ is the individual false positive rate of BFUs.
\label{fpTheorem}
\end{lemma}
\paragraph{Proof:} The probability of selecting a BFU which should return false is 
 $(1 - \frac{1}{B})$ if the multiplicity of the key is 1.
 If it is $V$ then the probability becomes $(1 - \frac{1}{B})^V$.
Since each Bloom Filter has a false positive rate $p$, the probability of introducing a false positive through a Bloom Filter failure is $ p(1 - \frac{1}{B})^V$. 
 
Because each BFU contains multiple documents, `True' documents (containing the query) can occur with `False' documents (not containing the query). Thus, we may also introduce false positives by merging $S_i$ into a BFU that contains the $k$-mer.
The probability for this event is
 $1 - (1 - \frac{1}{B})^V$.
The total per-document false positive rate for $R$ independent repetitions is $F_p = ( p(1 - \frac{1}{B})^V + 1 - (1 - \frac{1}{B})^V )^R $
Using this theorem, we can construct the overall false positive rate of RAMBO.

\begin{lemma}\textbf{ RAMBO False Positive Rate} \\
Given a RAMBO data structure with $B\times R$ BFUs, each with false positive rate $p$ and query $q$, we assume that $q$ belongs to no more than $V$ documents. Under this assumption, the probability of reporting an incorrect membership status for any of the $K$ documents, a.k.a. RAMBO False Positive Rate ($\delta$) is upper bounded by
$$\delta \leq K \left(1 - (1 - p)\left(1 - \frac{1}{B}\right)^V\right)^R $$ 
where $p$ is the individual false positive rate of the BFUs.
\label{overallFailureRate}
\end{lemma}
This is a direct result of Lemma \ref{fpTheorem} with union bound.
A consequence of lemma \ref{overallFailureRate} is that we need sub-linear RAMBO repetitions (logarithmic in $K$) to obtain an overall false positive rate $\delta$. We can state that it is sufficient to keep $R \geq {\log K - \log \delta}$. 
\begin{theorem} \textbf{Number of Repetitions} \\
\label{Rbound}
Given a set of $K$ files, maximum RAMBO false positive rate $\delta$ and $B$ Bloom Filter for each repetition, we need $R$ repetitions such that-
$$R = O(\log K -\log \delta)$$
\end{theorem}

\subsection{Query Time Analysis}
\label{queryTime}
This section demonstrates that RAMBO achieves sublinear query time in expectation. To query the set membership status of an element $x$, we perform $B \times R$ Bloom Filter look-ups followed by union and intersection operations (Section \ref{queryMethod}). 

Since each repetition makes a disjoint partition of the $K$ documents, the union operations do not require any computational overhead. The set intersections between repetitions, however, require $|X_1| + |X_2|$ operations, where $X_1$ is the set of all active documents in first repetition and $X_2$ is the set of all active documents in the next repetition. Since there are $R$ repetitions, the total cost for the intersection is $\sum_{r = 1 }^R |X_r|$. By observing that $\mathbb{E}[|X_r|] \leq V + Bp$, we obtain the following result. 

\begin{lemma} \textbf{Expected query time}\\
Given the RAMBO data structure with $B \times R$ BFUs and a query $q$ that is present in at most $V$ documents, the expected query time is
$$\mathbb{E}[q_t] \leq BR\eta+ \frac{K}{B}(V + Bp)R$$
where $K$ is the number of documents, $p$ is the BFU false positive rate, and $\eta$ is the number of hash functions used in BFUs. 
\end{lemma}

The first term represents the time to query the $B \times R$ BFUs. Note that $\eta$ ranges from $1$ to $6$ in practice. The second term is the time required to perform $R$ intersections.
We get $B = \sqrt{KV/\eta}$ by minimising the query time (i.e. solving $\nabla_{B} (\mathbb{E}[q_t])=0$). 
To obtain an expression for the query time in terms of the overall failure probability $\delta$ and the number of documents $K$, we suppose that $p \leq \frac{1}{B}$ and set $R$ according to Theorem~\ref{Rbound}. Our main theorem is a simplified version of this result where we omit lower-order terms. 

\begin{theorem} \textbf{RAMBO Query time}\\
Given a RAMBO data structure and a query $q$ that is present in at-most $V$ documents, RAMBO performs the search over $K$ documents with false positive rate $ \leq \delta$ in query time $q_t$, where
$$\mathbb{E}[q_t] = O\left(\sqrt{K} \left(\log{K} - \log {\delta}\right) \right)$$
Note that $V$ is independent of $K$ and $\delta$.
\label{MSMT_qt}
\end{theorem}

\subsection{Memory Analysis}
We provide an average case analysis under a simplifying assumption to analyze the expected performance of our method. We assume that every key has a fixed multiplicity $V$, meaning that every item is present in exactly $V$ documents. Under these assumptions, RAMBO requires the following amount of space.

\begin{table*}[ht]
\fontsize{9}{11}\selectfont
  \centering
  \begin{tabular}{ |c|c|c|c|c|c|c|c|c|c|c|c|c|c|c|c|c|c| } 
 \hline
 & \multicolumn{7}{|c|}{ Time per query (ms) (CPU time)} &    \multicolumn{5}{|c|}{ Construction time }  \\
 \hline
  & \multicolumn{4}{|c|}{ FASTQ} &  \multicolumn{3}{|c|}{ McCortex}  & \multicolumn{3}{|c|}{ FASTQ} & \multicolumn{2}{|c|}{ McCortex} \\
 \hline
\#files &  HowDe   &  SSBT  &  RAMBO  & RAMBO{$^+$} &  COBS &  RAMBO  & RAMBO{$^+$}&   HowDe      &  SSBT  &   RAMBO &  COBS  &  RAMBO \\
\hline 
100 &  5.24  & 8.47 & {\bf 0.018}& \textbf{0.0151} & 
0.19& {\bf 0.014}& \textbf{0.005} & 
2h30m & 52m    & 35m  & 
1m25s  & 1m12s\\ 
\hline 
200 & 10.38   & 17.12  & \textbf{0.025} & \textbf{0.0202}\  & 
0.38 & {\bf 0.017}& \textbf{0.011} & 
8h    & 1h47m     & 52m  &
3m58s & 2m25s\\  
\hline 
500 &  24.15  & 42.27  & \textbf{0.056} & \textbf{0.0483}\  &  
1.03 & {\bf 0.04}& \textbf{0.018} & 
21h   & 4h51m   & 1h57m &
8m28s & 6m22s\\ 
\hline
1000 & -  & 82.32 & \textbf{0.093} & \textbf{0.0747}\  & 
1.78 & {\bf 0.07}& \textbf{0.031} & 
-   & 9h16m    & 4h6m &
14m18s  & 12m32s\\
\hline
2000 & -  & 161.58 &\textbf{0.191} & \textbf{0.149}\   & 
2.72 & {\bf 0.09}& \textbf{0.059} & 
-  & 18h22m    & 8h55m  &
15m38s   & 25m41s\\
\hline
\end{tabular}
  \caption{Performance comparison between RAMBO and baselines on 1000 queries. To ensure a fair comparison, we have selected baseline hyper-parameters from their papers with the target false positive rate range of {$[0.01 ,0.011]$}, where the RAMBO false positive rate always falls in the range $[0.0095, 0.01]$. 
  HowDeSBT exceeds the available RAM on our platform after $500$ files.}
\label{HowDeSBTtable}
\vspace{-0.4cm}
\end{table*}

\begin{lemma} \textbf{Size of RAMBO} \\
\label{expectedMem}
For the proposed RAMBO architecture with size $B \times R$ and data with $K$ files, where every key has $V$ number of duplicates, the expected memory requirement is 
$$\mathbb{E}_v(M) = \Gamma \log K \log (1/p)\sum_{S \in \mathcal{S}}|S| \ \ \ \ \ \ \
\text{where} \ \Gamma <1 $$
\end{lemma}
Here $\Gamma = \sum^{V}_{v =1}\frac{1}{v}\frac{(B-1)^{V-2v+1}}{B^{V-1}}$.
The expectation is defined over the variable $v$ which takes values from \{1,2...V\}. This expression of $\Gamma $ holds if we are hashing document IDs using a universal hash function. If $B=K$, we will have one Bloom Filter per set. In that case, $\Gamma=1$.
We prove the expression of $\Gamma $ and its variation for any $B<K$ and $V>1$ in Section \ref{proofs}.
 
\section{Experiments}
\label{Experiments}
\subsection{Parameter Selection and Design Choices}
\label{paramSelection}
\noindent \textbf{Size of BFU:}
For each BFU to have a false positive rate $p$ using $\eta$ hashes, the size of the BFU must be set based on the number of insertions (Section \ref{BFintro}). One way to determine the BFU size is to preprocess the data by counting the number of terms that will fall into each BFU. 
In practice, it is sufficient to estimate the average set cardinality from a tiny fraction of the data, and we use this cardinality to set the size for all BFUs. Section  \ref{genomicExpe} presents these statistics for our data. \\ 
\noindent \textbf{B and R:}
 They are chosen according to $B = O(\sqrt{K})$ and $R = O(log K)$, where the constants were found empirically. \\
 \noindent \textbf{Bitmap arrays:}
 The intersection may be implemented using either bitmap arrays or sets. For binary operations, the OR operation is $O(1)$, hence intersection is very fast using bitmaps. The complexity of the AND operation depends on the set size; bitmaps are more efficient when 1s occupy > 15\% of the bitmap \cite{danielLemire}. This is true in our case, so we used bitmaps and found that the AND operations take fewer than $5\%$ of the query process cycles. Extensions such as SIMD-accelerated bitmaps are outside the scope of this work.\\
 \noindent \textbf{Query time speedup: } We may avoid querying all $B\times R$ BFUs by analyzing the repetitions sequentially. Specifically, in repetition $r$ we only need to query the BFUs that contain documents returned by previous repetitions $\{1,2...r-1\}$. BFUs that do not contain documents identified by previous repetitions cannot change the output, as any documents corresponding to those BFUs will be removed by the intersection operation of Algorithm~\ref{alg:membershipTest}.
We obtain a significant speedup using this sparse evaluation process (RAMBO$^+$ in Table \ref{HowDeSBTtable}).\\
 \noindent \textbf{k-mer size for sequence indexing: } The value of $k$ must be large enough that each gene sequence may be uniquely identified by a set of distinct $k$-mers. In the laboratory, the gene sequencing of an organism is done in parts by the sequencing machine. Here, each part (called "reads") is around $400-600$ in length typically. One might be tempted to set $k$ equal to the read length. However, portions of each read are often corrupted by errors, so a smaller $k$ must be used in practice. The work by Chikhi et.al. \cite{kmerSelection} confirmed k=31/51/71 to be optimal for $k$-mer set size and uniqueness. For the ENA dataset, most of the best methods \cite{bigSI} \cite{Bingmann2019COBSAC} \cite{SBT} \cite{HowDeSBT} use k=31, partially also because it is small enough to be represented as a 64-bit integer variable with 2-bit encoding. For these reasons and to provide a fair comparison with popular baselines, we use k=31. 
\begin{table}
\fontsize{9}{11}\selectfont
  \centering
  \begin{tabular}{ |c|c|c|c|c|c|c|c|c|c|c|c|c|c|c|c|c|c|c| } 
 \hline
 & \multicolumn{3}{|c|}{ Size (FASTQ)} &  \multicolumn{2}{|c|}{ Size (McCortex)}  \\
 \hline
\ \#files \ & \ HowDe\   & \ \ SSBT \ \ &  \ RAMBO\  & \ COBS \ &  \ RAMBO\ \ \\
\hline 
100 & 
92.5GB  & 9.5GB & 12.8GB &
2.4GB & 3.5GB \\ 
\hline 
200 & 
182GB  & 9.5GB & 19GB & 
4.9GB & 6.3GB \\ 
\hline 
500 & 
456GB  & 18GB & 42GB &  
7.5GB & 13.9GB \\ 
\hline
1000 & 
    -   & 36GB & 70GB &   
20GB& 23.2GB \\
\hline
2000 & 
-  & 72GB & 140GB & 
28GB& 47 GB \\
\hline
\end{tabular}
\caption{Size of index comparison for the same experiment from Table \ref{HowDeSBTtable}. In worst case, RAMBO takes $O(\log K)$ extra space than the optimal Array of Bloom Filter (COBS). HowDeSBT and SSBT uses RRR \cite{raman2007succinct} bitvector compression, however RAMBO does not compress the bitvectors. Any possible compression based optimization is left for the future exploration.}
\label{HowDeSBTtable2}
\vspace{-0.6cm}
\end{table}

\noindent\textbf{Dataset}: We use the 170TB WGS dataset (containing 460500 files) as described in~\cite{b3} and \cite{b12}. It is the set of all bacterial, viral, and parasitic gene sequence data in the European Nucleotide Archive (ENA) as of December 2016. The data is present in two formats - 1) FASTQ \cite{cock2010sanger} files containing raw, unfiltered sequence reads and 2) McCortex \cite{McCortex} \cite{bigSI} format, which is a filtered set of $k$-mers that omits low-frequency errors from the sequencing instruments. The $k$-mer length is 31 and the sequence alphabet (or nucleotides) are A, T, G, and C. Using 1000 random documents we found that the \{average, standard deviation\} number of $k$-mers is \{$377.6M$, $354.9M$\}, number of unique $k$-mers is \{$95M$, $103.1M$\} and file size is \{$145MB$, $86.5MB$\}. The high variation among documents demands either a preprocessing pass or our pooling method from Section \ref{paramSelection} to set the size of each BFU - we use the pooling procedure.

\vspace{-0.2cm}
\subsection{Genomic sequence indexing}
\label{genomicExpe}
We start our experiments by indexing only the first 2000 documents (2.4 TB). The results for the subset are shown in Table \ref{HowDeSBTtable}. The details are as follows
\\
\noindent\textbf{Baselines}: The COBS (Compact bit-sliced signature index)~\cite{Bingmann2019COBSAC} (Index based on an array of Bloom Filters) prefers McCortex data format and hence gives a very erroneous output on FASTQ. The Bloom Filter tree-based methods, SSBT ~\cite{b7} and HowDeSBT~\cite{HowDeSBT} works with FASTQ version but not with McCortex. Hence, we compare with COBS on McCortex and with SSBT and HowDeSBT on FASTQ.
The comparison with BIGSI is unnecessary, as COBS is the successor of BIGSI and is better in all aspects. The baseline implementations and RAMBO are in C++.

\noindent\textbf{Parameters}: For HowDeSBT, the Bloom Filter size is $7.5 \times 10^9$ bits. HowDeSBT only supports $1$ hash function and crashes if another value is used.
For SSBT, we use $4$ hash functions and set Bloom Filter size to $8.5 \times 10^8$ bits.
For COBS, we use $3$ hash function and set the false positive rate to 0.01. These parameters were hand-optimized and hard-coded into the program by the authors of COBS.
For RAMBO, we use $2$ hash functions, 
the number of partitions $B$ is $15, 27, 60, 100$ and $200$ for number of set insertions $100, 200, 500, 1000$ and $2000$. We set $R=2$ and the BFU size to $10^9$ bits for the McCortex data. For FASTQ, we use $R=3$ and $2 \time 10^9$ bits. 
These parameters are optimal for low query time, keeping in mind the allowable (comparable to baselines) index size, false positive rate, and construction time.\\
\noindent\textbf{Evaluation Metrics}: 
 Creating a test set with ground truth requires a very time-consuming procedure of generating inverted indices. 
The index's actual false positive rate can also be assessed by creating an artificial and unseen query-ground-truth set and inserting this set into the index before querying.
 Therefore, we calculated the false positive rate by creating a test set of $1000$ randomly generated $30$ length $k$-mer terms.
 We used length $30$ to ensure that there are no collisions from the existing $k$-mers already in the RAMBO data structure. These $k$-mers were assigned to $V$ files (distributed exponentially $(1/\alpha) \exp(-x/\alpha) $ with $\alpha=100$) randomly. The test set is much much smaller than the dataset's actual size; hence it makes an insignificant change in the size of RAMBO. 
 To get the index size, we report the maximum resident set size (RSS) in memory as returned by the \textit{time} utility or the serialized index size, whichever is higher. This size includes the main index as well as all auxiliary data structures (like the inverted index mapping $B$ buckets to $K$ documents). Query time is the CPU time on a single thread, and construction time is the wall-clock time on 40 threads (with no other process running on the machine).

\noindent\textbf{System and Platform Details}: We ran the experiment on a cluster with multiple $40$ core Intel(R) Xeon(R) Gold 6230 CPU @ 2.10GHz processor nodes. Each node has 192 GB of RAM and 960 GB of disk space. The experiments, apart from RAMBO construction on the full dataset, are performed on a single node. We did not use multi-threading for querying.
 
  From Table \ref{HowDeSBTtable} we can see that RAMBO has much faster query time (from around \textbf{25x} to \textbf{2000x}) than the baselines. Furthermore, RAMBO achieves a small index size (practically close to the theoretical lower bound - the array of Bloom Filters). The construction of RAMBO is an I/O bound process; hence we see almost linear growth in construction time (with number of files), which is equivalent to COBS and faster than SSBT and HowDeSBT. Insertion from McCortex format is blazing fast and preferred as it has unique and filtered $k$-mers.
\vspace{-0.2cm}
\subsection{Smart parallelism- Indexing the full 170TB WGS dataset in 9 hours from scratch}
\label{smartPar}
We now address the construction of RAMBO for the entire 170TB dataset. One could theoretically create a single RAMBO data structure with the given R and B parameters on a single machine. However, this is infeasible due to the limited DRAM and compute resources. We could also construct the index over multiple machines using a message passing interface, but this introduces a massive latency overhead due to data transmission over the network. A third way is using a shared-memory cluster of machines, which requires a massive DRAM and is infeasible in current multi-core servers.

We propose a better solution. We parallelize the computation by partitioning the RAMBO data structure over 100 nodes of the cluster. Each node contains a small RAMBO data structure indexing $1/100$ of the whole dataset, which is around $4605$ files in our case. In the streaming setting, a file (set of terms) is routed to a BFU of a node randomly. 
More details about routing are the following-

\noindent\textbf{Routing:}
We first use a random hash function $ \tau (.)$ to assign files to node and then use an independent smaller node-local 2-universal hash function $\phi_i(.)$ to assign the file to the local Bloom Filter (BFU). This process preserves all the mathematical properties and randomness in RAMBO as the final mapping is again 2-universal, i.e., the probability of any two datasets colliding is exactly $1/B$, where $B$ is the total range (number of partitions in RAMBO).
The two-level hash function is given by: $(b \times \tau(D_j)) + \phi_i(D_j)$.
For a repetition $i$, where $i \in \{0..R\}$, $b$ is the number of partitions in RAMBO on a single machine and also the range of $\phi(.)$, $D_j$ is the name ID of $j^{th}$ dataset, and the range of $\tau(.)$ is $\{0..100\}$ in our case. Note that this two-level hash function allows us to divide the insertion process into multiple disjoint parts (100 in our case) without repeating any installation of datasets and internode communications. Effectively, each node will contain a set of $4605$ files in expectation.
In this way, we eliminate costly transmission of data among the nodes. The data structure on each node has size $B =500$ and $R=5$. Stacking them vertically makes the complete RAMBO data structure of size $B =50000$ and $R=5$. 
This process preserves the randomness of set insertion,  i.e., the probability of any two sets colliding is exactly $1/B$, where $B$ is the total range ($50000$ in this case). 

\begin{figure}[ht]
\centering
\includegraphics[scale = 0.47]{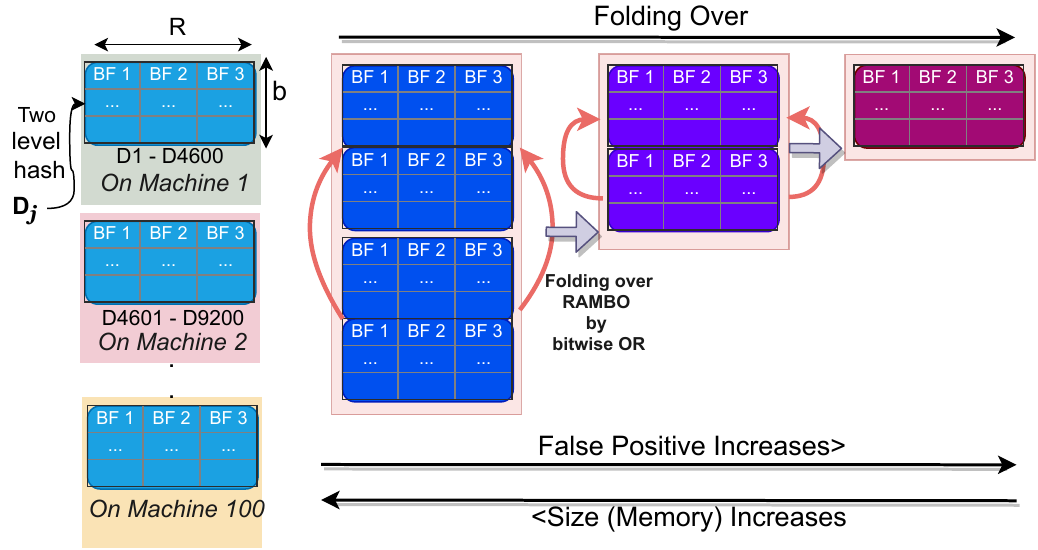}
\caption{Indexing process of 460K documents over a cluster of 100 nodes. Each machine carries a part of RAMBO with size $500 \times 5$ Bloom Filters. The dataset is routed to machine via $\tau(.)$ hash functions followed by insertion using $\phi_i(.)$. The combined direct routing is done by a two-level hash function equivalent ($b*\tau(.) + \phi(.)$). The stacked view of RAMBO shows the folding process. The folding is done such that number of repetitions $R$ remains the same but $B$ halves, so as the total size. Folding reduces memory progressively by factors of 2, 4, 8... and increases false positive rate super-linearly.}
\label{fig:foldingfig}
\end{figure}


\begin{table}[ht]
\fontsize{9}{11}\selectfont
    \centering
    \begin{tabular}[b]{|c|c|c|}\hline
      & Query Time (CPU time in ms)   &  Index  size   \\
  \hline 
  Fold 2 & 66.5 & 7.13 TB  \\ 
 \hline
  Fold 4 & 43.5  & 3.6 TB  \\ 
 \hline 
  Fold 8 & 26.25  & 1.78 TB \\ 
 \hline 
    \end{tabular}
    \caption{CPU time (in ms) per query of the k-mer averaged over 1000 queries. Each column shows the different number of RAMBO folds. Second column shows the memory size (in TB) of RAMBO for each fold.}
    \label{tab:my_label}
\end{table} 

\vspace{-0.6cm}

\begin{figure}[ht]
    \includegraphics[scale=0.4]{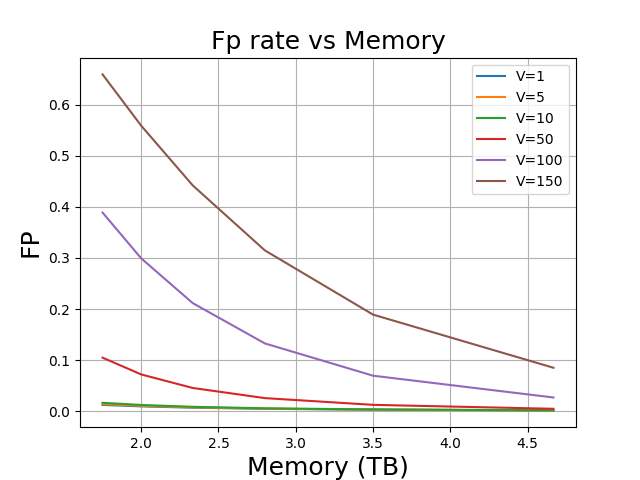}
    \vspace{-0.4cm}
    \caption{False positive rate of RAMBO for different values of V (k-mer multiplicity per $4605$ sets) and memory. Note that the false positive rates are very low if query is rare. For a full sequence search, the returned result depends solely on the rarest k-mer. Hence our method returns very accurate (low false positives) results. }
    \label{fig:6}
  \end{figure}

This interesting parallel insertion trick results in a fully constructed RAMBO in \textbf{about an hour on 100 CPU nodes when using the McCortex file format}. The additional 8 hours are used to download the dataset. It is the round-off time of the highest time taking job. Here we have to ensure that all machines use the same parameters ($B, R$, Bloom Filter size and hash function $\tau(.)$, $\phi(.)$ and $h(.)$) as well as the random seeds. The consistency of seeds across machines and larger than required $B$ and $R$ allow us to flexibly reduce the size of RAMBO later by doing bitwise OR between the corresponding BFUs of the first half of RAMBO over the other half (vertically). Each of these processes reduces the index size $B \times R$ to $\frac{B}{2} \times R$. This is called folding over. Refer Figure \ref{fig:foldingfig}.\\
\noindent\textbf{Folding Over:} The data structures on every machine are independent and disjoint, but they have the same parameters and uses the same hash seeds. Since RAMBO is all made of Bloom Filters, we can perform the bit-wise OR between the first half of RAMBO over the other half to reduce the partitions in RAMBO from $B$ to $\frac{B}{2}$. This operation is depicted in Figure~\ref{fig:foldingfig}. With this folding-over, a one-time processing allows us to create several versions of RAMBO with varying sizes and FP rates (Table \ref{tab:my_label} and Figure \ref{fig:6}).
\vspace{-0.4cm}
  
\vspace{-0.3cm}
\begin{table}[h]
\fontsize{9}{11}\selectfont
  \centering
  \begin{tabular}{ |c|c|c|c|c|c|c|c|c|c|c|c|c|c|c|c|c|c|c|c|c| } 
 \hline
 & \multicolumn{3}{|c|}{ Wiki-dump (17K)} &  \multicolumn{3}{|c|}{ClueWeb (50K)}  \\
 \hline 
 &  QT(ms)  &  Size  &  CT  &  QT(ms)  &  Size  &  CT   \\
\hline 
HowDe  & 3.781 & 6.43GB & 101m & 1.5 & 8GB & 5h\\
\hline 
COBS  & 0.523 & 157MB & 2.71s & \textbf{0.56} & 88M &7.6s\\ 
\hline
RAMBO &\textbf{ 0.074}  &  \textbf{51 MB} & \textbf{1.75s} & 0.58 &\textbf{62M} & \textbf{5.3s} \\
\hline 
\end{tabular}
  \caption{ Performance comparison between RAMBO and baselines on wiki-dump data and part ClueWeb data on false positive rate of $0.01$. QT is time per query (CPU time) in ms.}
\label{WebDoc}
\end{table}
\vspace{-0.8cm}

\subsection{Document indexing}
We extend our experiments for web data where each document is represented as a set of English words.\\
\noindent\textbf{Datasets}: We use a sample from Wiki-dump \cite{wiki-dump} and the popular TREC Category B ClueWeb09 dataset\cite{clueweb}. The Wiki-dump sample has 17618 documents 
The ClueWeb09 dataset sample has 50K (non-spam) documents of the English language. Both datasets were pre-processed by removing stop words, keeping only alpha-numeric, and tokenizing as word unigrams. Wiki dump is $207$ MB, and Clueweb is $98MB$ after pre-processing.

\noindent\textbf{Parameters}: 
For Wiki dump, RAMBO has $B=1000, R =2$, and the size of each BFU is $200000$ bits. ClueWeb09, we choose $B=5000, R =3$, and size of each BFU = $20000$ bits. Clueweb has shorter files (~$450$ terms per file) than Wiki-dump (~$650$ terms per file).

\noindent\textbf{Baseline}: We compare with the COBS and Sequence Bloom Tree as in Section \ref{Experiments}.

\noindent\textbf{Evaluation Metric}: 
We created a query set of randomly generated terms other than what is present in the data. We inserted them using an exponentially distributed term multiplicity $V$, similar to the experiment on genomic data.
We perform experiments on the same system as in section \ref{genomicExpe}. The query is performed sequentially on a single core and thread for a fair comparison. Refer Table \ref{WebDoc}.

\section{Discussion}
RAMBO provides a solid trade-off between false positive rate and query time while retaining all desirable properties of Bloom Filter and the bitsliced data structure. Due to cheap updates, RAMBO takes very little time for index creation (Table \ref{tab:my_label} and Section \ref{smartPar}).
RAMBO performs updates on the stream and is embarrassingly parallel for both insertion and query. The false positive rate of RAMBO is very low for low term multiplicity (Figure \ref{fig:6}). This low false positive rate is guaranteed for full sequence/phrase queries, as the rarest of the terms dominates. Therefore, RAMBO can perform a quick and accurate check of an unknown and rare gene sequence. 
Furthermore, due to sublinear scaling, RAMBO becomes more efficient in memory at a large scale. This property will allow RAMBO to be used as an efficient search engine for extreme-scale applications. 

\vspace{-0.2cm}
\section{Appendix}
\label{proofs}

\textbf{Lemma 4.6 Proof}: 
We want to find the unique insertions in each $B$ Bloom Filter and sum them up to get the size of a single table in RAMBO. If $B=1$, the unique insertions will be $\frac{N}{V}$ where $N = \sum_{S \in \mathcal{S}}{|S|}$ is the total number of insertions. If we partition these documents into $B$ bins, every term from the dataset has varying number of duplicates $v$ where $v \in \{0,1,2...V\}$ in a bin. 0 duplicate implies that the term does not exist in the given bin, and 1 duplicate implies that the term has only one copy. Note that each bin corresponds to a BFU where the terms/kmers are inserted.
The expected number of unique terms going in bin $b$ is given by: 
$|b| = \mathbb {E}\Big[\sum_{i}^{N_b}\frac{1}{v}\Big]$, where there are $N_b$ is the number of insertions in $b^{th}$ bucket and $\frac{1}{v}$ is a random variable, $\frac{1}{v} \in\{{1, \frac{1}{2}, \frac{1}{3}, .....\frac{1}{V}}\}$. 
By the linearity of expectation, we can state that
$$|b| = \sum_{i}^{N_b}\mathbb {E}\Big[\frac{1}{v}\Big] = \sum_{i}^{N_b}\sum^{V}_{v =1}\frac{1}{v} \times p_v$$
We can view $\frac{1}{v}$ as a multiplicity reduction factor of a term.
Here, $P_v$ is the probability of getting $v$ balls in one bucket and $V-v$ in remaining others. Hence we can write 
$P_v =  \frac{1}{B^{v-1}} \times \Big(\frac{B-1}{B}\Big)^{V-v} $.
This gives the expected size of all the bins in a table-
$$ \sum_{i}^{N}\sum^{V}_{v =1}\frac{1}{v}\frac{(B-1)^{V-2v+1}}{B^{V-1}} = \sum_{S \in \mathcal{S}}|S| \sum^{V}_{v =1}\frac{1}{v}\frac{(B-1)^{V-2v+1}}{B^{V-1}} $$
 As $B<K$ and $V>1$, $\Gamma <1$ always, where $ \Gamma = \sum^{V}_{v =1}\frac{1}{v}\frac{(B-1)^{V-2v+1}}{B^{V-1}} $\\
The expected size of RAMBO is given by
$$ \Gamma \log K \log (1/p)\sum_{S \in \mathcal{S}}|S|$$

\bibliographystyle{ACM-Reference-Format}
\balance
\bibliography{template}

\end{document}